\documentclass{article}

\usepackage{PRIMEarxiv}

\usepackage[utf8]{inputenc} 
\usepackage[T1]{fontenc}    
\usepackage{hyperref}       
\usepackage{url}            
\usepackage{booktabs}       
\usepackage{amsfonts}       
\usepackage{nicefrac}       
\usepackage{microtype}      
\usepackage{lipsum}
\usepackage{fancyhdr}       
\usepackage{graphicx}       
\graphicspath{{media/}}     

\usepackage{amsmath,amssymb,amsfonts}
\usepackage{textcomp}
\usepackage{xcolor}
\usepackage{braket}
\usepackage{tikz}
\usetikzlibrary{quantikz2}
\usepackage{stfloats}
\usepackage{makecell}
\usepackage{xcolor}
\usepackage{mathtools}
\usepackage{paralist}
\usepackage{subcaption}
\usepackage{pifont}
\usepackage[most,breakable]{tcolorbox}
\usepackage{xspace}
\usepackage{bbding}
\usepackage{amsthm}
\usepackage{colortbl}
\usepackage{tabularx}
\usepackage{siunitx}
\usepackage{nicefrac}
\usepackage[inline]{enumitem}
\usepackage{hyperref}
\usepackage{graphicx}
\usepackage{array}
\usepackage{booktabs}
\hypersetup{
colorlinks = true, 
hidelinks,
linkcolor=black, 
citecolor=black, 
urlcolor=black 
}

\newcommand{\ourApproach}{\texttt{QuReBot}\xspace}
\newcommand{\QRConly}{\texttt{QRC-only}\xspace}
\newcommand{\Skiponly}{\texttt{Skip-only}\xspace}
\newcommand{\Atwelve}{\ensuremath{\hat{A}}\textsubscript{12}\xspace}
\newcommand{\LMSE}{\ensuremath{\mathcal{L}_{\mathit{MSE}}}\xspace}
\newcommand{\FS}{\ensuremath{\mathit{FS}}\xspace}
\newcommand{\horizon}{\ensuremath{K}\xspace}

\title{Quantum Machine Learning-based Test Oracle for Autonomous Mobile Robots
}

\author{
   Xinyi Wang \\
  Simula Research Laboratory and \\
  University of Oslo \\
  Oslo, Norway\\
  \texttt{xinyi@simula.no} \\
   \And
   Qinghua Xu \\
  Lero Research Centre and \\
  University of Limerick \\
  Limerick, Ireland\\
  \texttt{qinghua.xu@ul.ie} \\
  \AND
  Paolo Arcaini \\
  National Institute of Informatics \\
  Tokyo, Japan \\
  \texttt{arcaini@nii.ac.jp} \\
  \And
  Shaukat Ali \\
  Simula Research Laboratory and \\
  Oslo Metropolitan University\\
  Oslo, Norway \\
  \texttt{shaukat@simula.no} \\
  \And
  Thomas Peyrucain \\
  PAL Robotics \\
  Barcelona, Spain \\
  \texttt{thomas.peyrucain@pal-robotics.com} \\
}

\begin{document}
\maketitle

\begin{abstract}
Robots are increasingly becoming part of our daily lives, interacting with both the environment and humans to perform their tasks. The software of such robots often undergoes upgrades, for example, to add new functionalities, fix bugs, or delete obsolete functionalities. As a result, regression testing of robot software becomes necessary. However, determining the expected correct behavior of robots (i.e., a test oracle) is challenging due to the potentially unknown environments in which the robots must operate. To address this challenge, machine learning (ML)-based test oracles present a viable solution. This paper reports on the development of a test oracle to support regression testing of autonomous mobile robots built by PAL Robotics (Spain), using quantum machine learning (QML), which enables faster training and the construction of more precise test oracles. Specifically, we propose a hybrid framework, \ourApproach, that combines both quantum reservoir computing (QRC) and a simple neural network, inspired by residual connection, to predict the expected behavior of a robot. Results show that QRC alone fails to converge in our case, yielding high prediction error. In contrast, \ourApproach converges and achieves 15\% reduction of prediction error compared to the classical neural network baseline. Finally, we further examine \ourApproach under different configurations and offer practical guidance on optimal settings to support future robot software testing.
\end{abstract}


\section{Introduction}

\textit{PAL Robotics}~\cite{pal} is a European company developing robots to improve daily life and solve practical problems. Their robots support domestic and industrial tasks using self- and environmental awareness for safe autonomous navigation with built-in path planners. This paper focuses on {\it Autonomous Mobile Robots} (AMRs), which are widely deployed in service environments such as offices, where they operate in cooperation with humans. During navigation with data collected from onboard sensors, AMRs frequently interact with the environment, including humans and other objects; this can result in failures if AMRs are not thoroughly tested.

As with other software systems, robot software continuously evolves through updates and enhancements to improve performance. These changes require regression testing to ensure that existing, updated, and newly introduced functions all operate correctly. However, the oracle problem remains a challenge in regression testing, as obtaining ground truth is often difficult~\cite{he2019system, arrieta2021using, AitorPaper}. In some cases, it can even be very costly or even impossible, especially when additional equipment and complex setup are required~\cite{afzal2020study}. In addition, simulating robots to build ground truth costs substantial time and resources. To this end, machine learning (ML)-based test oracles have been employed in the literature~\cite{fontes2021using, arrieta2021using, AitorPaper} as a viable solution to learn expected behavior from historical data and predict correct expected behavior during testing. 

This work develops an ML-based test oracle to support AMR regression testing. It specifically targets the navigation software that enables AMRs to follow an optimal path to their destination while avoiding collisions with humans or other obstacles. Traditional ML models require training on substantial amount of data, so necessitating significant computational resources. In contrast, {\it reservoir computing} is a computational framework that uses the natural dynamics of randomly connected neural networks to process information, with its main advantage being minimal training requirements~\cite{butcher2013reservoir}. The neural networks used in reservoir computing are typically {\it recurrent neural networks} (RNNs), which are well-suited for capturing temporal dependencies in sequential data~\cite{tanaka2019recent}.


Recent progress in {\it quantum computing} (QC) has triggered interest in {\it quantum reservoir computing} (QRC)~\cite{mujal2021opportunities}, which replaces the RNN-based reservoir with a QC system. 
By combining QC's computational power with the minimal training capability of reservoir computing, QRC offers the potential for highly efficient data processing. In QC, classical data is mapped into high-dimensional Hilbert spaces, enabling the model to better capture temporal patterns and make accurate predictions, which, in our context, means a more accurate test oracle learned with minimal training.

Thus, in this work, we propose a QRC-based model as a test oracle for the navigation software of AMRs developed by \textit{PAL robotics}. Specifically, we train this model to perform next state prediction of the AMRs based on a sequence of historical robot states. The model is trained on data from a stable software version and later used to test the behavior of updated versions, ensuring existing functionality remains unaffected.

Although QRC has shown promise in various domains, such as trajectory prediction, quantum chemistry problems, and gene regulatory networks~\cite{fujii2017harnessing, PhysRevResearch.3.013077, martinez2020information, domingo2022optimal, domingo2023taking, mlika2023user}, most existing QRC approaches are still limited to handling univariate time series data. In contrast, multivariate time series data (e.g., AMR state data), despite extensively-studied in traditional ML, remains underexplored in QRC. This gap motivates us to combine QRC with traditional ML approaches to effectively and efficiently tackle complex multivariate time series data.


To that end, we propose a hybrid framework, \ourApproach, which is inspired by the {\it residual connection} technique in deep learning~\cite{he2016deep}. It has been widely applied in several deep learning algorithms, such as ResNet~\cite{wu2019wider}, and Transformer architectures~\cite{vaswani2017attention} to reduce the influence of vanishing gradient by passing the input directly to the output layer, skipping the intermediate layers. In our context, \ourApproach uses QRC for time series data prediction and a residual connection to link the input to the final output layer through a relatively simple non-linear transformation. For each input data, \ourApproach dynamically calculates a weight parameter to control the contribution of the QRC branch and the residual connection branch (the ``shortcut'') based on its context.

We evaluate the performance of \ourApproach with the AMR developed by \textit{PAL robotics}.\footnote{All experiment results and code for replication can be found at~\cite{qurebot}.} Results show that QRC alone fails to converge and results in high prediction error. \ourApproach converges successfully and achieves a 15\% lower MSE compared to the classical baseline. Finally, we compare \ourApproach across various configurations and provide practical recommendations for optimal settings for future robot development.

\section{Background}
\subsection{Quantum Computing}\label{sec:qc}
The quantum computing paradigm leverages quantum mechanical principles to solve certain problems that are difficult for classical computers. It performs computations using {\it quantum bits} (i.e., \textit{qubits}), instead of bits, to process information. Unlike a bit, which can only be either 0 or 1, a qubit can exist in a \textit{superposition} of state $\ket{0}$ and $\ket{1}$ at the same time, until it is \textit{measured} and collapses into one of the basis states with different probabilities. The \textit{quantum state} (i.e., also called as {\it pure state}) of a single qubit can be represented as
\begin{equation}
\ket{\psi} = \alpha_0\ket{0} + \alpha_1\ket{1} =
\begin{bmatrix}
\alpha_0 \\
\alpha_1
\end{bmatrix}, \quad |\alpha_0|^2 + |\alpha_1|^2=1
\end{equation}
where the probability of the qubit being $\ket{0}$ is $|\alpha_0|^2$ while that of being $\ket{1}$ is $|\alpha_1|^2$. The quantum state of a multi-qubit system with $N$ qubits can be represented by
\begin{equation}
\ket{\psi} = \sum_{i=0}^{2^N - 1} \alpha_i \ket{i} =
\begin{bmatrix}
\alpha_0 \\
\alpha_1 \\
\vdots \\
\alpha_{2^N - 1}
\end{bmatrix}, \quad \sum_{i=0}^{2^N - 1} |\alpha_i|^2 = 1
\end{equation}
where $\ket{i}$ denotes the $i$-th computational basis state in binary.

Another important concept in quantum computing is \textit{entanglement}, where two or more qubits share a single joint quantum state and depend on each other. For example, in the entangled two-qubit quantum state:
\begin{equation}
\ket{\psi} = \frac{1}{\sqrt{2}}\big( \ket{00} + \ket{11} \big)
\end{equation}
the system collapses into either $\ket{00}$ and $\ket{11}$ with equal probability when they are measured.

\begin{figure}[!tb]
\centering
\includegraphics[width=0.4\linewidth]{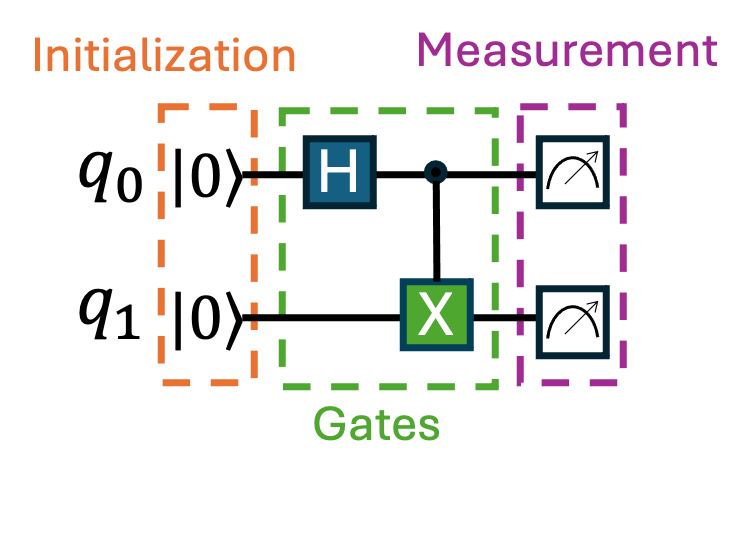}
\caption{Example of quantum circuit}
\label{fig:circuit}
\end{figure}

{\it Gate-based quantum computing} is a widely used model to perform computations by applying a sequence of \textit{gates} (i.e., \textit{unitary operators}) to a \textit{quantum circuit}, which is composed of several qubits. Table~\ref{table:gatetype} shows some commonly used gates.
\begin{table}[!tb]
\caption{Descriptions of commonly used quantum gates}
\label{table:gatetype}
\footnotesize
\resizebox{\columnwidth}{!}{
\begin{tabular}
{m{0.19\columnwidth}|m{0.8\columnwidth}}
\toprule
\textbf{Gate} & \textbf{Description} \\
\midrule
Hardamard \textit{(H)} & Placing a qubit into an equal superposition, i.e., there is a 50\% probability of being measured in state either $\ket{0}$ or $\ket{1}$.\\ \hline
\textit{Pauli-Z} (\textit{Z}) & Rotating a qubit around the $z$-axis with $\pi$ radians. \\
\hline
\textit{Pauli-X} (\textit{X}) & Rotating a qubit around the $x$-axis with $\pi$ radians. \\
\hline
\textit{Pauli-Y} (\textit{Y}) & Rotating a qubit around the $y$-axis with $\pi$ radians. \\
\hline
$RX(\theta)$ & Rotating a qubit around the $x$-axis with $\theta$ radians. \\
\hline
$RY(\theta)$ & Rotating a qubit around the $y$-axis with $\theta$ radians.\\
\hline
$RZ(\theta)$ & Rotating a qubit around the $z$-axis with $\theta$ radians. \\
\hline
\textit{Controlled-X (CNOT/CX)} & A two-qubit gate with a control and a target qubit. If the control qubit is $\ket{1}$, the target qubit rotates around the $x$-axis with $\pi$ radians. \\
\hline
\textit{Controlled-Z (CZ)} & A two-qubit gate with a control and a target qubit. If the control qubit is $\ket{1}$, the target qubit rotates around the $z$-axis with $\pi$ radians. \\
\hline
Two qubit rotation gates \textit{RXX/RYY/RZZ} & A two-qubit gate rotating the two qubits around $z$ axes simultaneously with $\theta$ radians.\\
\bottomrule
\end{tabular}
}
\end{table}
In a quantum circuit, each qubit is first initialized into a classical basis state (e.g., $\ket{0}$), then quantum gates are applied to manipulate their quantum states (e.g., creating superposition and entanglement) to perform the desired computation. For example, Fig.~\ref{fig:circuit} shows a two-qubit circuit, where the two qubits are first initialized in the $\ket{00}$ state.
An $H$ gate is applied on the $q_0$, placing it into superposition. Next, a $\mathit{CX}$ gate entangles the two qubits. Finally, the two qubits are measured, causing the system to collapse into a classical basis state.

\subsection{Quantum Reservoir Computing (QRC)}\label{sec:qrc}
\noindent\textbf{Classical Reservoir Computing.} {\it Reservoir Computing} (RC) is a computational framework for temporal information processing. RC leverages a fixed, non-linear system called \emph{reservoir} to map the input sequence into higher-dimensional spaces, similar to neural network hidden layers. Then, a \emph{readout} layer is employed to produce the desired output sequence~\cite{tanaka2019recent}. Unlike neural networks where all weights are trained, RC keeps the \emph{reservoir} fixed and only trains the \emph{readout} layer. Consequently, RC is computationally more efficient and requires less data for training compared to neural networks. 

\begin{figure}[!tb]
\centering
\includegraphics[width=0.8\linewidth]{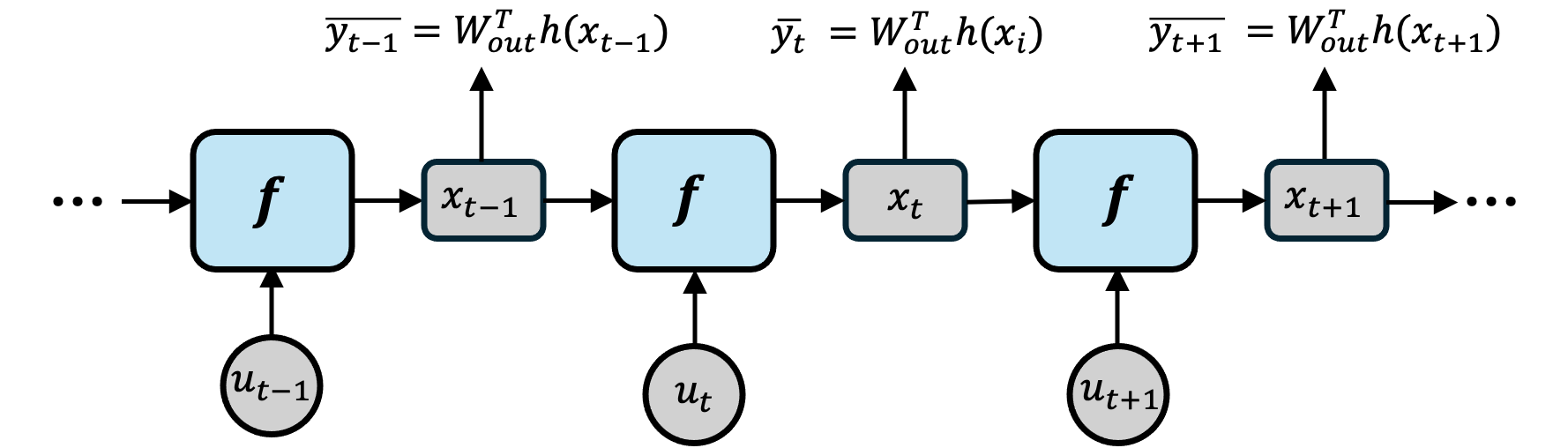}
\caption{Framework of Reservoir Computing (RC)}
\label{fig:rc_framework}
\end{figure}

Specifically, given an input sequence $\{u_t\}^{M-1}_{t=0}$, RC aims to predict the corresponding output sequence $\{y_t\}^{M-1}_{t=0}$. Fig.~\ref{fig:rc_framework} shows the general framework of RC.
To encode the temporal information of sequences, the \emph{reservoir} in RC maintains an internal state $x_t$, referred to as the \emph{reservoir state} hereafter. The update rule of $x_t$ at timestep $t$ is given by
\begin{equation}\label{eq:internal}
x_t = f(x_{t-1}, u_t)
\end{equation}
where $f$ is a non-linear, fixed function that represents the temporal evolution of the reservoir state. Similar to the recurrent unit of RNN, $f$ incorporates both information from the previous state $x_{t-1}$ and the current input $u_t$ to predict the current state $x_t$. Subsequently, RC maps $x_t$ to the output space through a \emph{readout} layer, which is essentially a linear transformation defined as
%
\begin{equation}\label{eq:output}
\bar{y_t} = {W_\mathit{out}}h(x_t)
\end{equation}
where ${W_\mathit{out}}$ is a weight matrix trained to reduce the difference between the predicted output $\bar{y_t}$ and the target output $y_t$. $h$ is a function that observes signals from the state $x_t$.

\begin{figure}[!tb]
\centering
\includegraphics[width=0.8\linewidth]{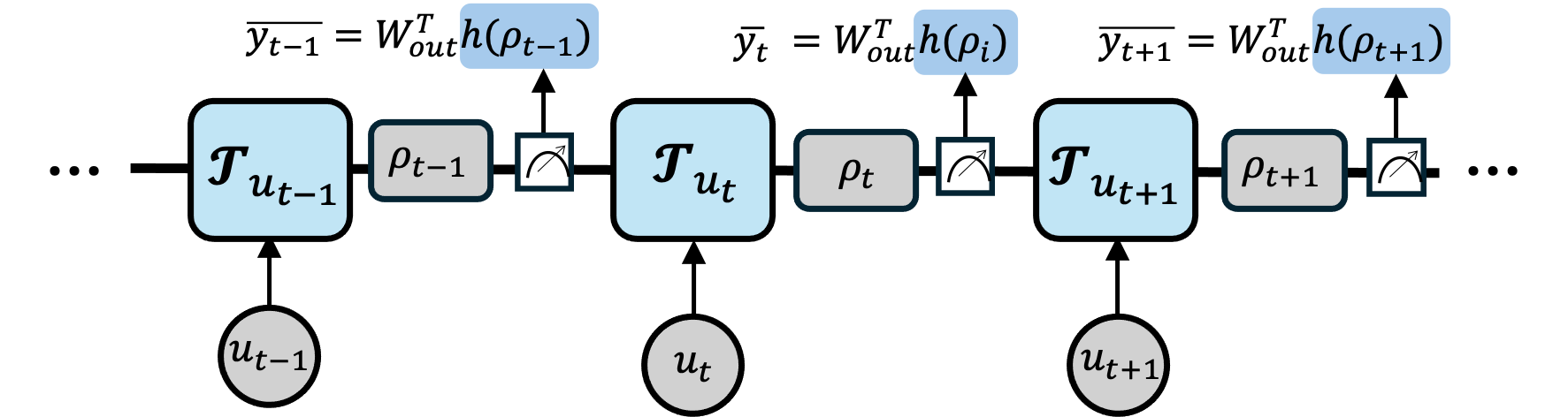}
\caption{Framework of Quantum Reservoir Computing (QRC)}
\label{fig:qrc_framework}
\end{figure}

\begin{figure}[!tb]
\centering
\includegraphics[width=0.4\linewidth]{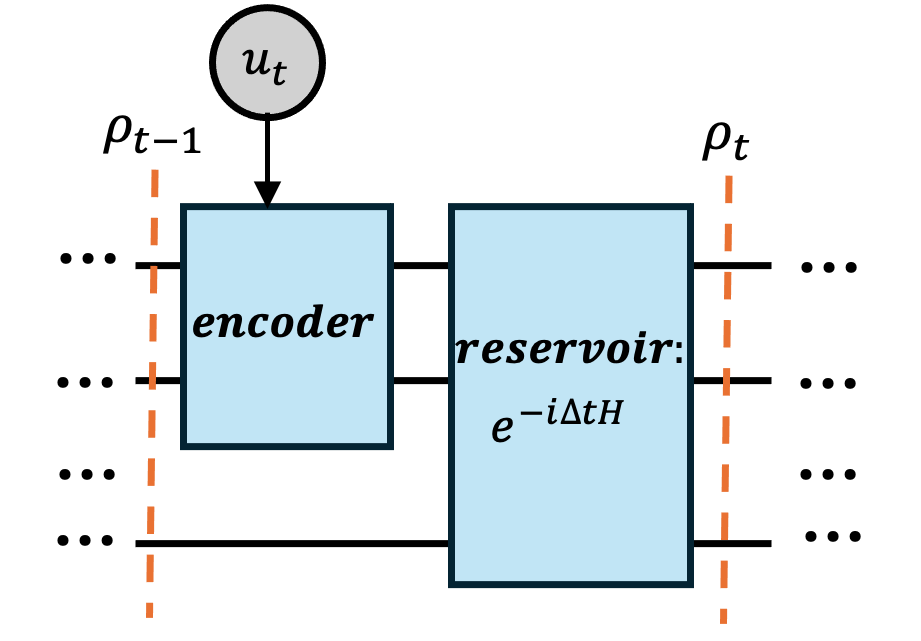}
\caption{$\mathcal{T}_{u_t}$ Circuit}
\label{fig:qrc_unit}
\end{figure}

\noindent\textbf{Quantum Reservoir Computing}. {\it Quantum Reservoir Computing} (QRC) extends the classical reservoir in RC with a quantum system as depicted in Fig.~\ref{fig:qrc_framework}.
%
Following the general framework of RC, we inject values of input sequence $\{u_t\}^{M-1}_{t=0}$ at successive timesteps into the quantum circuit to calculate corresponding internal \emph{reservoir states} (i.e., quantum states in this context), which are then used to predict the output values through a \emph{readout} layer. As discussed in Sect.~\ref{sec:qc}, the (pure) quantum state of a quantum system with $N$ qubits is represented as a $2^N$-dimensional vector $\ket{\psi}$. Since a QRC system can be considered as a statistical mixture of pure states, we represent it as a ($2^N \times 2^N$)-dimensional density matrix $\rho$. The time evolution of the quantum system under an input $u_t$ is described by
\begin{equation}
\rho_t=\mathcal{T}_{u_t}(\rho_{t-1})=e^{-i\Delta t H}\rho(t-1)e^{i\Delta t H}
\end{equation}
where $\rho_t$ represents the internal reservoir state at time $t$. $\mathcal{T}_{u_t}$ represents the evolution process in Fig.~\ref{fig:qrc_unit} where the input value $u_t$ is first encoded and applied to the quantum system, which evolves for time $\Delta t$ under the unitary operator $e^{-iH\Delta t}$ defined by the system Hamiltonian $H$.\footnote{Hamiltonian defines the energy of a quantum system, and is mathematically expressed as a hermitian matrix.}
Notably, $H$ is configurable, and it decides the system dynamics. During evolution, the injected information will spread through the whole system. The function $h$ in Eq.~\ref{eq:output} is implemented by measurements on all qubits. We measure the system and calculate the expectation values on each qubit, obtaining
\begin{equation}
h(\rho_t)=[\langle Z_0 \rangle, \ldots, \langle Z_{N-1} \rangle]
\end{equation}
After getting sufficient pairs of $\big(h(\rho_t), y_t\big)$, we can train a simple linear regression model to determine the parameters of $W_{\mathit{out}}$ by minimizing the mean squared error between predicted outputs $\bar{y_t}$ and target outputs $y_t$:
\begin{equation}
\min \sum_{t=0}^{M}(\bar{y_t}-y_t)^2 
\end{equation}

\begin{figure}[!tb]
\centering
\includegraphics[width=0.98\linewidth]{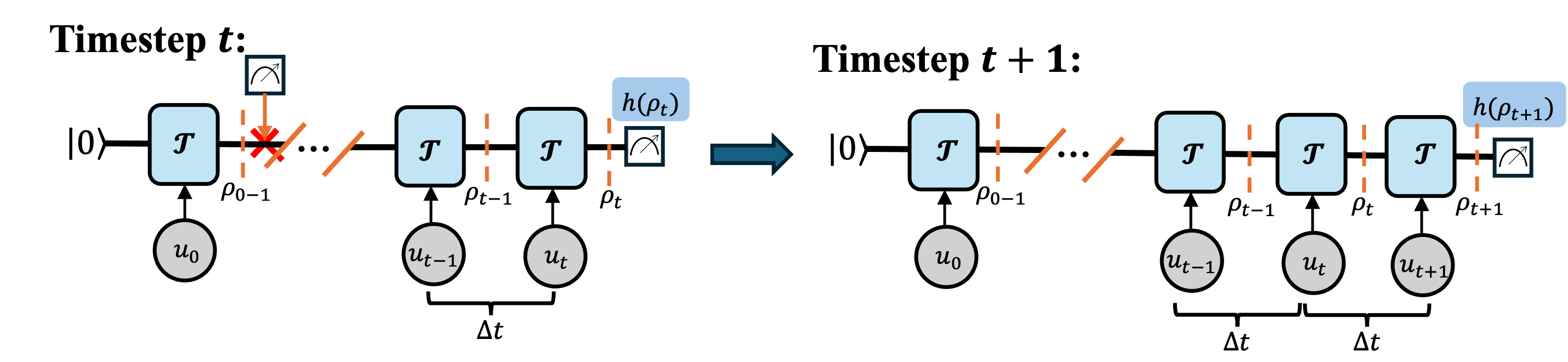}
\caption{Restarting Protocol}
\label{fig:restart}
\end{figure}

However, during implementation, the computational cost of this process is considerable. First, due to the probabilistic nature of the quantum system, multiple repetitions are necessary to approximate expectation values accurately. Second, the above framework requires measurements on all qubits between every two timesteps to get $h(\rho_t)$, which is destructive as it causes the whole system to collapse and disrupt the system dynamics. Consequently, the entire process must be repeated from the beginning for each timestep.
For example, as illustrated in Fig.~\ref{fig:restart}, to obtain $h(\rho_t)$ at timestep $t$, the system evolves from $u_0$ to $u_t$ before measurement.
Then, to get $h(\rho_{t+1})$, the system is reset and evolved again from $u_0$ to $u_{t+1}$. This \textit{restarting protocol} is highly inefficient and requires substantial resources.

\begin{figure}[!tb]
\centering
\includegraphics[width=0.98\linewidth]{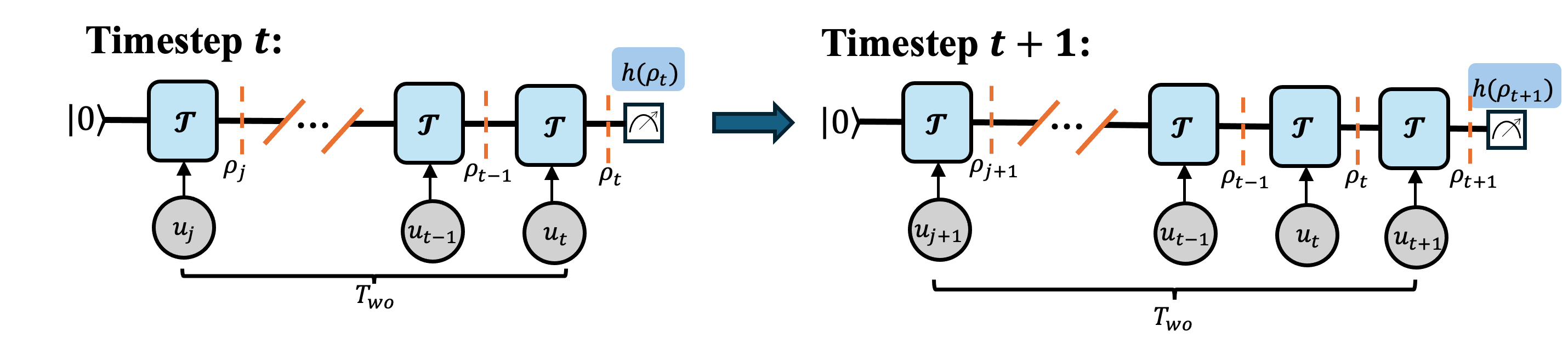}
\caption{Rewinding Protocol}
\label{fig:rewind}
\end{figure}

To address this issue, a \textit{rewinding protocol} is proposed based on the {\it echo state property} in RC, which illustrates that, after sufficient evolution, reservoir state becomes independent of its initial state, depending only on recent inputs~\cite{grigoryeva2018echo}. Accordingly, in implementation, we introduce the washout time $T_{\mathit{wo}}$ as the timesteps necessary for RC to ``forget'' the initial state. Thus, as shown in Fig.~\ref{fig:rewind}, instead of restarting from $t_0$ every time, the system only needs to reconstruct the last $T_{\mathit{wo}}$ steps, which significantly reduces the required resources.

\section{Industrial Context}

PAL Robotics is a Spanish company with world-leading expertise in developing robots for deployment in the service industry (e.g., warehouses and retail). They support businesses and are also actively involved in research and development projects with research organizations in this domain. Examples of robots built by PAL include legged robots for research purposes, robots designed to assist in retail environments (e.g., for stocking goods), and Autonomous Mobile Robots (AMRs), which are widely used in various logistics applications~\cite{pal}.

Although the work presented in this paper can be applied to various types of robots produced by PAL, we here focus on AMRs, specifically the TIAGo OMNI robot~\cite{tiago_omni_base}. This robot is widely used in both research and industrial settings for its accurate navigation and advanced sensing capabilities.
It is commonly deployed in service environments such as office spaces and hospitals, where it operates alongside humans. Consequently, it must navigate in a way that ensures human safety, while effectively completing its assigned tasks. The robot employs two navigation algorithms: the first one is designed to determine an optimal route to its destination, while the second one focuses on detecting obstacles and dynamically adapting the robot's path. This adaptive behavior is supported by real-time data from multiple onboard sensors that continuously monitor the surrounding environment.

Like any other software system, a robot's software evolves, requiring regression testing to ensure that the updated software continues to meet both its functional and non-functional requirements. Such testing requires a test oracle to determine whether the robot's behavior remains correct, i.e., navigation remains optimal and safe in our context. However, building a test oracle is challenging, as the ground truth is often complicated to obtain. To address this challenge, machine learning-based test oracles are often employed~\cite{fontes2021using, arrieta2021using, AitorPaper}, typically trained on historical data. In line with this approach, we investigate the use of QML to develop a test oracle that supports the regression testing of the TIAGo OMNI robot. Specifically, we design a hybrid framework that integrates neural networks and QRC. The quantum dynamics within QRC enhances the neural networks' ability to learn more precise patterns in data compared to classical machine learning. 
As a result, the hybrid model can model the navigation behavior of the robot more precisely and accurately, thereby serving as a precise test oracle to support regression testing.



\section{Approach}
We introduce the overview of \ourApproach. First, we describe how we integrate QC with a classical neural network to predict the next states of a robot based on its previous states. Second, we elaborate on how we design the core of \ourApproach, i.e., the quantum reservoir circuit.

\subsection{Overview}\label{sec:overview}
\begin{figure*}[!tb]
\centering
\includegraphics[width=1\linewidth]{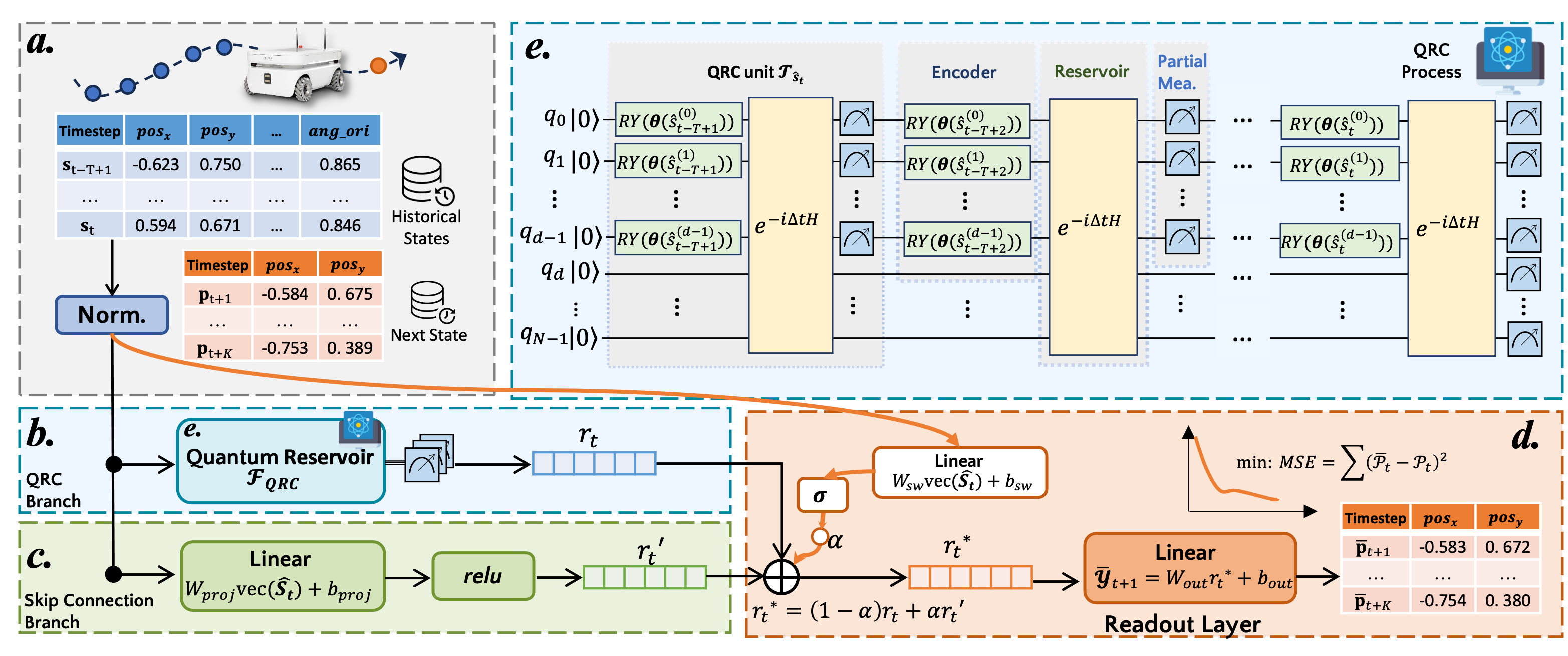}
\caption{Overview of \ourApproach}
\label{fig:overview}
\end{figure*}

Fig.~\ref{fig:overview} shows an overview of the training process in \ourApproach.
For training, we gather navigation data during operation from a stable release of the TIAGo OMNI robot. As shown in Block $a$ of Fig.~\ref{fig:overview}, we consider a sequence of $t+1$ observed historical state $\mathcal{S}=\{ \mathbf{s}_0,$ $\mathbf{s}_1,$ $\dots, \mathbf{s}_t\}$, where each state contains features related to its position, orientation, linear velocity, and angular velocity. Specifically, each state consists of $d$ features as in $\mathbf{s}_j=\{s_j^{(0)},$ $s_j^{(1)},$ $\ldots,$ $s_j^{(d-1)}\}$. Considering the rewinding protocol illustrated in Sect.~\ref{sec:qrc}, we select features of the last $T$ states (i.e., washout time) as the input of our model (i.e., $\mathcal{S}_t=\{ \mathbf{s}_{t-T+1},$ $\mathbf{s}_{t-T+2},$ $\ldots,$ $\mathbf{s}_t\}$). We also define the number of future time steps to predict as horizon \horizon. The output of the model is the position of the robot in the next \horizon time steps 
$\mathcal{P}_{t}=\{\mathbf{p}_{t+1},$ $\mathbf{p}_{t+2},$ $\ldots,$ $\mathbf{p}_{t+\horizon}\}$ and $\mathbf{p}_{j} = (p_j^{\mathit{pos}_x}, p_j^{\mathit{pos}_y})$.

\ourApproach is a QRC-based approach to predict the robot's position in the next \horizon states. Inspired by the success of skip connections in deep neural networks, we adopt a similar strategy in \ourApproach to enhance prediction effectiveness. In deep learning, skip connections typically link the input directly to the output layer, bypassing intermediate layers. The adoption of skip connections can potentially preserve critical information in the input, mitigating severe context loss in deep neural networks or complex quantum reservoirs. As shown in Fig.~\ref{fig:overview}, our approach contains two branches: the {\it QRC branch} (i.e., Block $b$) and the {\it skip connection branch} (i.e., Block $c$).

In the QRC branch, to process the input data into quantum circuits, we first normalize each state as $\mathbf{\hat{s}}_j=\{\hat{s}_j^{(0)},$ $\hat{s}_j^{(1)},$ $\ldots,$ $\hat{s}_j^{(d-1)}\}$ to fit the requirements of the quantum encoders. The normalized data then undergoes time evolution within the reservoir circuits. After measurement, we get expectation values for each qubit, i.e., a real-valued vector defined as:
\begin{equation}
\mathbf{r}_t = \mathcal{F}_{\mathit{QRC}}(\mathbf{\hat{\mathcal{S}}}_t) = \{r_t^{(0)}, r_j^{(1)}, \ldots, r_j^{(D-1)}\}
\end{equation}
where $\mathcal{F}$ is the QRC process. More details in Sect.~\ref{sec:qrc_impl}.

The skip connection branch provides a ``shortcut'' to bypass the complex quantum reservoir, consisting of relatively simple transformations. As depicted in Block \textit{b} of Fig.~\ref{fig:overview}, the normalized state $\hat{s}_j$ is consecutively fed into a linear layer and an activation layer (ReLU), yielding a vector $\mathbf{r}_t'$ with the same dimension as $\mathbf{r}_t$. The transformation equations are as follows
%
%
%
%
\begin{align}
& \mathbf{r}_t'= W_{\mathit{proj}} \cdot \operatorname{vec}(\mathcal{\hat{\mathcal{S}}}_t) + b_{\mathit{proj}} \, , \quad \operatorname{vec}(\mathcal{\hat{S}}_t) \in \mathbb{R}^{Td}\\
& \mathbf{r}_t' = relu(\mathbf{r}_t')
\end{align}
where $W_{\mathit{proj}}$ and $b_{\mathit{proj}}$ are weight matrices.
%

As shown in Block \textit{d} of Fig.~\ref{fig:overview}, \ourApproach calculates a weighted sum of both outputs from the QRC branch ($\mathbf{r}_t'$) and ($\mathbf{r}_t$). Standard residual connections add $\mathbf{r}_t'$ and $\mathbf{r}_t$ directly, implicitly treating both inputs equally, whereas \ourApproach dynamically adjust the influence of each branche based on the characteristics of the input. Specifically, \ourApproach computes a scalar value $\alpha$ from the input $\hat{\mathcal{S}}_t$ as 
%
%
\begin{equation}
\label{eq:alpha}
\alpha = W_{\mathit{sw}} \cdot \operatorname{vec} (\hat{\mathcal{S}}_t) + b_{\mathit{sw}}
\end{equation}

Then, a \textit{sigmoid function} $\sigma$ is applied to map the weight into a value between 0 and 1.
\begin{equation}
\alpha=\sigma(\alpha)
\end{equation}

With $\alpha$ computed, \ourApproach computes the weighted sum $\mathbf{r}_t^*$ of $\mathbf{r}_t'$ and $\mathbf{r}_t$ as in 
\begin{equation}
\mathbf{r}_t^* = (1-\alpha) \mathbf{r_t} + \alpha \mathbf{r}_t'
\end{equation}

Finally, a readout layer takes $\mathbf{r}_t^*$ as input and predicts the target output. Specifically, the readout layer is a linear transformation defined as
\begin{equation}
\mathcal{\bar{P}}_{t}=W_{\mathit{out}} \cdot \mathbf{r}_t^* + b_{\mathit{out}}
\end{equation}
$\mathcal{\bar{P}}_{t}$ represents the predicted sequence of next \horizon time states. The objective of this hybrid model is to minimize the mean squared error between the predicted output $\mathcal{\bar{P}}_{t}$ and the target output $\mathcal{P}_{t}$. Suppose $M$ input-output pairs are given, the objective is formally formulated as
\begin{equation}
\min \operatorname{MSE}=\frac{1}{M}\sum_{t=0}^{M-1}(\mathcal{\bar{P}}_{t}-\mathcal{P}_t)^2
\end{equation}

\subsection{Quantum Reservoir Computing Implementation}\label{sec:qrc_impl}
In this section, we introduce the QRC process $\mathcal{F}_{\mathit{QRC}}$ (i.e., Block $e$ in Fig.~\ref{fig:overview}). The first issue we need to address is to feed the sequence of the input states $\hat{\mathcal{S}}_t$ into the quantum system. In QRC, each state is injected into the quantum circuit at consecutive time steps. At each time step, each normalized feature value $\hat{s}_t^{(j)} \in [0, 1]$ is fed to the system by setting the quantum state of one qubit to $\ket{\psi_{\hat{s}_t^{(j)}}}=\sqrt{1-\hat{s}_t^{(j)}}\ket{0}+\sqrt{\hat{s}_t^{(j)}}\ket{1}$. To set this quantum state, we need to apply an $\mathit{RY}$ gate (see Table~\ref{table:gatetype}) on the $j$-qubit with rotation $\theta=2\arcsin \sqrt{\hat{s}_t^{(j)}}$. As shown in the \textit{Encoder} box in Block $e$ of Fig.~\ref{fig:overview}, we select a subset of qubits to encode the feature values. Each qubit is encoded with one value.

Then, all qubits in the quantum system undergo time evolution through an input-independent unitary operator as the reservoir circuit, which is commonly a variational quantum circuit. The parameters in the reservoir circuit are randomly assigned and fixed throughout the process. Various structures for reservoir circuits have been proposed. Due to the existing hardware restrictions on current quantum computers, a class of hardware-efficient~\cite{tang2021qubit} quantum reservoir circuits has been proposed. These circuits are constructed from repeated layers. Each layer includes an optional {\it rotation layer}, consisting of single-qubit rotation gates, and an {\it entangling layer}, built from multi-qubit gates. The number of repetitions of these layers determines the circuit \textit{depth}. In this paper, we consider the following four quantum circuits as quantum reservoir circuits.

\begin{figure}[!tb]
\centering
\begin{subfigure}[b]{0.4\columnwidth}
\centering
\resizebox{\columnwidth}{!}{
\begin{quantikz}
\lstick{$q_0$} & \ctrl{1} & & \gate{X} & \\
\lstick{$q_1$} & \gate{X} & \ctrl{1} & & \\
\lstick{$q_2$} & & \gate{X} & \ctrl{-2} & \\
\end{quantikz}
}
\caption{\textit{CNOT} Reservoir}
\label{fig:cnot_circuit}
\end{subfigure}
\begin{subfigure}[b]{0.4\columnwidth}
\centering
\resizebox{\columnwidth}{!}{
\begin{quantikz}
\lstick{$q_0$} & \gate{RY(\theta_0)} & \ctrl{1} & & \ctrl{2} & \\
\lstick{$q_1$} & \gate{RX(\theta_1)} & \ctrl{-1} & \ctrl{1} & & \\
\lstick{$q_2$} & \gate{RZ(\theta_2)} & & \ctrl{-1} & \ctrl{-2} & \\
\end{quantikz}
}
\caption{Rotation Reservoir}
\label{fig:rotation_circuit}
\end{subfigure}

\vspace{0.5em}

\begin{subfigure}[b]{0.46\columnwidth}
\centering
\resizebox{\columnwidth}{!}{
\begin{quantikz}
\lstick{$q_0$} & \gate{RY(\theta_0)} & \gate{RZ(\theta_3)} & & \ctrl{1} & \\
\lstick{$q_1$} & \gate{RY(\theta_1)} & \gate{RZ(\theta_4)} & \ctrl{1} & \gate{X} & \\
\lstick{$q_2$} & \gate{RY(\theta_2)} & \gate{RZ(\theta_5)} & \gate{X} & & \\
\end{quantikz}
}
\caption{Efficient SU2 Reservoir}
\label{fig:effsu2_circuit}
\end{subfigure}
\begin{subfigure}[b]{0.48\columnwidth}
\centering
\resizebox{\columnwidth}{!}{
\begin{quantikz}
\lstick{$q_0$} & \gate{RX(a_0)} & \gate[2]{RZZ(J_{0,1})} & \qw & \gate[3]{RZZ(J_{1,2})} & \qw \\
\lstick{$q_1$} & \gate{RX(a_1)} & & \gate[2]{RZZ(J_{1,2})} & & \qw \\
\lstick{$q_2$} & \gate{RX(a_2)} & \qw & & & \qw \\
\end{quantikz}
}
\caption{Ising Hamiltonian Reservoir}
\label{fig:ising_circuit}
\end{subfigure}
\caption{Quantum Reservoir Circuits}
\label{fig:reservoir_circuits}
\end{figure}
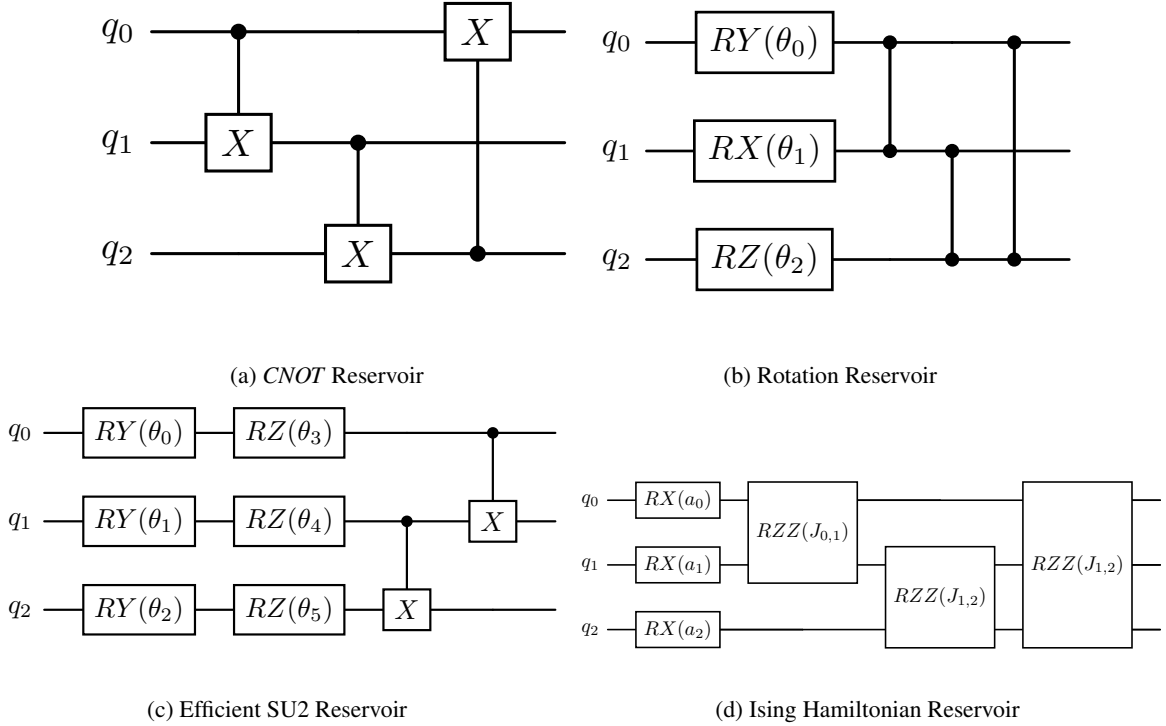

\vspace{1em}
\begin{compactitem}
\item[\textbf{(1) CNOT Reservoir:}] It does not include a rotation layer and contains no randomly assigned parameters. It is built only from $\mathit{CX}$ gates, that are arranged in a circular structure to entangle all the qubits. Fig.~\ref{fig:cnot_circuit} shows an example circuit with three qubits in a single repetition.

\item[\textbf{(2) Rotation Reservoir:}] It contains a rotation layer and an entangling layer in a single repetition. The rotation layer applies a single-qubit rotation gate on each qubit, randomly selected from the gate set $\{\mathit{RX},$ $\mathit{RY},$ $\mathit{RZ}\}$. Each rotation angle $\theta_i$ is randomly sampled from the range $[0, 2\pi)$ and fixed. The entangling layer is constructed by $\mathit{CZ}$ gates arranged in a circular entangling structure to connect all qubits. Fig.~\ref{fig:rotation_circuit} shows an example circuit with three qubits in a single repetition.

\item[\textbf{(3) Efficient SU(2) Reservoir:}] It is based on a widely used ansatz in Qiskit for variational quantum algorithms~\cite{twolocal}. Each repetition consists of a rotation layer followed by an entangling layer. In the rotation layer, we apply two single-qubit rotation gates on each qubit, which are $\mathit{RY}$ and $\mathit{RZ}$ gates. The rotation angles $\theta_i$ are randomly selected from $[0, 2\pi)$ and remain fixed throughout the process. The entangling layer is constructed by $\mathit{CX}$ gates arranged in a ``reverse linear'' entangling structure, where every two adjacent qubits are entangled starting from the higher-indexed qubit. Fig.~\ref{fig:effsu2_circuit} shows an example circuit with three qubits in a single repetition.

\item[\textbf{(4) Ising Hamiltonian Reservoir:}] It corresponds to Ising Hamiltonian dynamics:
\begin{equation}
H = \sum_{j=1} a_j X_j + \sum_{j<k} J_{jk} Z_j Z_k
\end{equation}
where $a_j$ is called the {\it local field coefficient} specifying how strongly qubit $j$ interacts with the field along the $X$ axis. $J_{jk}$ shows the {\it coupling coefficient} between qubit $j$ and $k$, which describes the interaction between the two qubits along $Z$ axis. Thus, in the rotation, we apply an $RX$ gate on each qubit with parameter $a_j$ as the rotation angle. In the entangling layer, $RZZ$ gates are applied to connect qubits in a circular structure, with rotation angle $J_{jk}$. The values of both $a_j$ and $J_{jk}$ are randomly sampled from the range $[-1, 1]$. Fig.~\ref{fig:ising_circuit} shows an example circuit with three qubits in a single repetition.
\end{compactitem}
\vspace{1em}
%

In this model, we select a reservoir circuit with a set of fixed parameters to train the model. The entangling layer ensures the inputs injected into the circuit spread throughout the system. As shown in Sect.~\ref{sec:qrc}, each QRC unit $\mathcal{T}_{s_t}$ contains an encoder and a reservoir circuit. Although we apply the rewinding protocol, constructing a QRC circuit of $T$ time steps leads to a long circuit, which is challenging as quantum systems can lose coherence quickly, making them unreliable. To address this, partial measurements (i.e., instead of full measurement) between every two timesteps are performed to balance memory and practicality~\cite{yasuda2023quantum}. Specifically, we measure the qubits where inputs are injected while leaving the remaining \textit{ancilla qubits} unmeasured so they can continue evolving and preserve past input information. The measured qubits will get reset to inject new input.

After the system evolves $T$ time steps, we measure all the qubits at $Z$ basis to obtain the final vector $\mathbf{r}_t=\{r_t^{(0)},$ $r_j^{(1)},$ $\ldots,$ $r_j^{(D-1)}\}$, which also includes information collected from partial measurements during previous timesteps.

\section{Experiment Design}\label{sec:expDesign}
\subsection{Research Questions}
\begin{compactitem}
\item[\textbf{RQ0.}] {\it How is the performance of QRC with different quantum reservoirs}? 

We conducted a pilot study to evaluate the performance of using only the QRC algorithm for predicting the robot's next state. In addition, we aim to assess the impact of different quantum reservoir circuits on the performance.

\item[\textbf{RQ1.}] {\it How is the performance of \ourApproach compared to the classical baseline model}?

This RQ compares the performance of \ourApproach and a simple classical neural network model. We also compare the two models under various configurations to evaluate our model's robustness.

\item[\textbf{RQ2.}] {\it What is the performance of \ourApproach across different input features and prediction horizons}?

This RQ evaluates the performance of \ourApproach under different configurations. Based on the results, we provide practical guidelines of optimal input features and prediction horizons to use for researchers and practitioners using \ourApproach.
\end{compactitem}

\subsection{Datasets and Features}
We employed the open-source PAL Robotic OMNI Base Simulation environment~\cite{pal_robotics_simulator} to create the dataset, which integrates ROS~2 and Gazebo to provide a realistic 3D simulation of the TIAGo OMNI Base robot. The robot navigated through a set of waypoints in the PAL Robotics office environment using the ROS~2 Nav2 stack. We recorded a total of 14,236 time steps at a rate of 10 Hz frequency, to train our model. Each time step captured seven features in three categories position (i.e., $\mathit{pos}_x$, $\mathit{pos}_x$), orientation (i.e., $\mathit{ori}_z$, $\mathit{ori}_w$), and velocity (i.e., $\mathit{vel}_x$, $\mathit{vel}_y$, ${\mathit{{vel}_{ang}}}$). To investigate the effect of feature selection on \ourApproach, we create three feature sets to train our model: $\FS_7$ (all features), $\FS_5$ (without orientation), and $\FS_4$ (without velocity). Notice that we do not consider the exclusion of position-related features, as they are indispensable for our task, i.e., predicting the next position of the robot.

\subsection{Baselines}
We consider two baselines: (1) 
\QRConly: This is the baseline QRC model without enhancements. In contrast to \ourApproach, this model removes the skip connection branch. The inputs undergo the quantum evolution, and the final outputs are predicted based solely on the measurement of the quantum circuit with a linear regression; (2) \Skiponly: This is a pure classical model with a simple neural network. It differs from \ourApproach by removing the QRC branch. The inputs are passed through a linear layer, a \textit{ReLU} activation function, and another linear layer to predict the outputs.

\subsection{Parameter settings}

\textbf{QRC circuit configuration:} To reduce the influence of randomness in the reservoir, we fix a single random seed to generate the variational circuits, which are then consistently used across all QRC components. To balance performance and computational cost, we set the depth for the quantum reservoir circuit to 10. Also, we use one extra qubit as the ancilla qubit. 

\textbf{Experiment configuration:} For \QRConly, we did a pilot study and implemented QRC with all four reservoir circuits described in Sect.~\ref{sec:qrc_impl}. We configure the model with feature set $\FS_7$ and horizon $\horizon=1$. For \ourApproach, we use the Ising Hamiltonian reservoir in the QRC branch. For both \Skiponly and \ourApproach models, we implement all three feature sets (i.e., $\FS_7$, $\FS_5$, $\FS_4$) with prediction horizons ranging from 1 to 5.

\textbf{Training parameters:} To fully utilize the dataset, we conduct 4-fold cross-validation by splitting the dataset into four equal subsets based on the time sequence. Each experiment uses one fold for testing and the remaining three for training. Moreover, we select 20\% of the training datasets as the validation datasets. The model is trained with mean squared error as the loss function, with a learning rate of $1 \times 10^{-4}$ and a batch size of 32. Training runs for up to 500 epochs, with early stopping applied if the validation loss does not improve for 10 consecutive epochs. The final model is selected based on the lowest validation loss. To mitigate randomness, each experiment is repeated 10 times. As a result, a total of $5 \times 3 \times 4 \times 10 = 600$ models were trained for both \ourApproach and \Skiponly, respectively, for evaluation. \looseness=-1


\subsection{Experiment setup}
We implement the QRC component of \ourApproach with the \textit{Quantumreservoirpy} package~\cite{Kulseng2024quantumreservoirpy}, based on the Qiskit framework. We use the Qiskit Aer simulator to execute quantum circuits. The quantum computing process is run on AMD EPYC Naples 7601 (SMT2) w/2TB of RAM and 4TB of NVMe scratch. \ourApproach and the baselines are trained on AMD EPYC Milan 7763 64-core w/ 8 qty Nvidia Volta A100/80GB.

\subsection{Evaluation metrics and statistical tests}
To evaluate the effectiveness of \ourApproach and the baselines, we calculate the average \textit{mean squared error} as:
\begin{equation}
\LMSE = \frac{1}{|\mathit{TS}| \cdot \horizon \cdot |\operatorname{\mathbf{p}}| }\sum_{ts=1}^{|\mathit{TS}|}\sum_{j=1}^{\horizon}(\operatorname{\mathbf{p}}_{\mathit{ts}, j}-\operatorname{\bar{\mathbf{p}}}_{\mathit{ts}, j})^2
\end{equation}
where $|TS|$ represents the size of the test dataset, \horizon represents the horizon in prediction, and $\operatorname{\bar{\mathbf{p}}}_{\mathit{ts}, j}$ refers to the predicted vector of the output (i.e., the position).

In RQ0, for each reservoir circuit, we gather \LMSE of \QRConly model across 4 folds and 10 repetitions for each reservoir circuit. To assess whether there are significant performance differences among various reservoir circuits, we apply the Kruskal-Wallis test. A $p$-value less than 0.05 indicates a statistically significant difference among the circuits; otherwise, no significant difference is observed.

In RQ1, to implement \ourApproach and the \Skiponly model under various configurations, for each configuration we calculate \LMSE across 4 folds and 10 repetitions. To compare the performance of the two models, we use the Mann-Whitney U test combined with \Atwelve effective size. If the yielded $p$-value is less than 0.05, it means that there is a significant difference in performance between the two models. In this case, we calculate the \Atwelve statistics. If the \Atwelve value is less than 0.5, it shows that \ourApproach outperforms the \Skiponly model, while a value above 0.5 indicates the opposite.

In RQ2, we compare the overall performance of \ourApproach with different feature sets and horizons respectively. Considering the feature sets, we group \ourApproach with the same feature set (but varying prediction horizons) and compare these groups. We first use the Friedman test to determine if there are significant differences among all three groups. A $p$-value less than 0.05 indicates statistically significant differences. Then, we perform pairwise comparisons using the Wilcoxon signed rank test with \Atwelve effect size. If $p$-value is smaller than 0.05, it shows there is a significant difference between the pair. Then, the \Atwelve statistic is calculated to determine the strength of the statistical test result. If the \Atwelve value is less than 0.5, it shows that \ourApproach with the first feature set is likely to be better than the second one, and vice versa. The same statistical procedures are applied to compare \ourApproach models across different prediction horizons.

\section{Results and analyses}
\subsection{RQ0: Performance of \QRConly with different reservoirs}\label{sec:rq0}
We evaluate the performance of the \QRConly model using the same parameters as in training \ourApproach. However, during training, validation loss failed to converge within 500 epochs, and predictions deviated significantly from actual values. This may be due to the nature of the input data: robot states recorded at fine-grained time steps show minimal variation between consecutive states. When a highly dynamic quantum reservoir processes such low-variation input, the measured output may not effectively capture subtle transitions. Moreover, the multivariate nature of the time series further complicates the task of identifying meaningful patterns. Without additional context, the linear readout layer struggles to learn meaningful patterns to predict output accurately.

This behavior was consistently observed on the \QRConly model with all four reservoir circuits. We computed \LMSE using 4-fold cross-validation with 10 repetitions. The average \LMSE over 40 runs is around 15.6 for all four reservoir circuits, indicating similar prediction error across them.
%
Statistical tests show that the $p$-value is larger than 0.05, indicating that there is no significant difference among those structures.

\begin{tcolorbox}[size=title, colframe=white, width=1\linewidth,
breakable,
colback=gray!20]
\textbf{Answer to RQ0}:
\QRConly fails to converge, resulting in high prediction error. In addition, using different reservoir circuits in \QRConly has no significant influence on its predictive effectiveness in our case.
\end{tcolorbox}

\subsection{RQ1: Performance of \ourApproach comparing to \Skiponly model}\label{sec:rq1}
We compare \ourApproach with the \Skiponly model to evaluate the contribution of the QRC branch in enhancing the prediction accuracy. Based on the findings in RQ0 (Sect.~\ref{sec:rq0}), we observe that the structure of the reservoir circuit has minimal impact in our scenario. Thus, we implement \ourApproach using the widely adopted Ising Hamiltonian Reservoir~\cite{kutvonen2020optimizing, gotting2023exploring}. In addition, since we evaluate \ourApproach in multi-step prediction tasks, we vary the prediction horizon \horizon from 1 to 5. Moreover, we conduct feature ablation experiments by using different sets of features to analyze if \ourApproach can preserve performance through reservoir dynamics despite a reduced number of features.

\begin{figure}[!tb]
\centering
\begin{subfigure}[b]{0.16\columnwidth}
\includegraphics[width=\columnwidth]{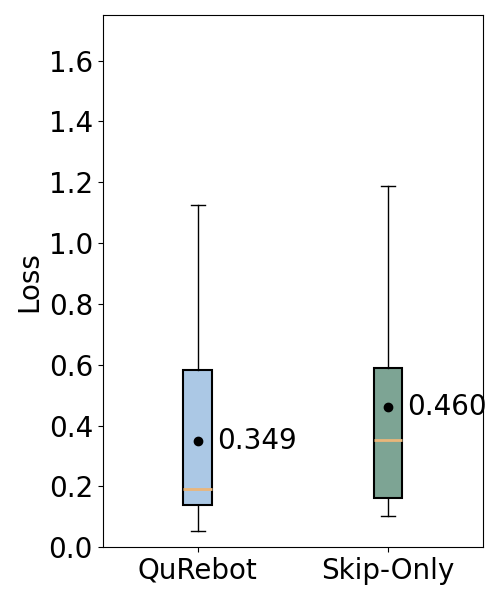}
\caption{$\FS_7$, $\horizon=1$}
\end{subfigure}
\begin{subfigure}[b]{0.16\columnwidth}
\includegraphics[width=\columnwidth]{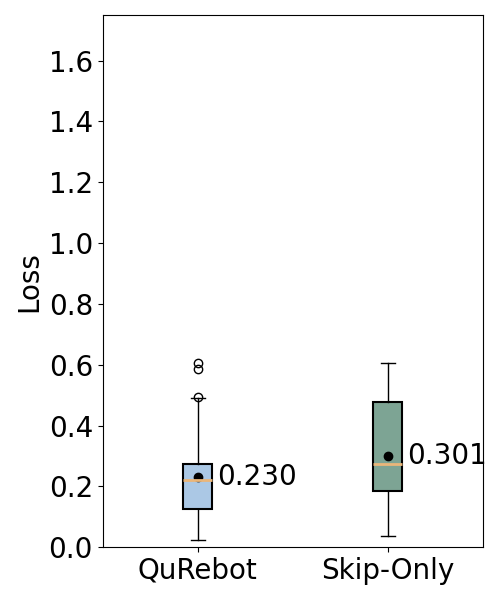}
\caption{$\FS_5$, $\horizon=1$}
\end{subfigure}
\begin{subfigure}[b]{0.16\columnwidth}
\includegraphics[width=\columnwidth]{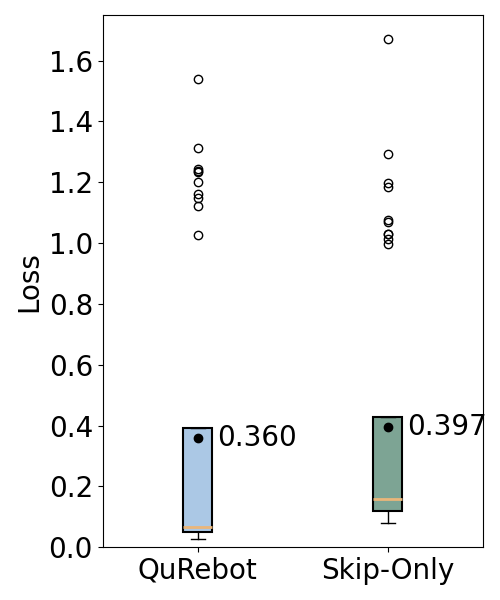}
\caption{$\FS_4$, $\horizon=1$}
\end{subfigure}

\vspace{0.5em}

\begin{subfigure}[b]{0.16\columnwidth}
\includegraphics[width=\columnwidth]{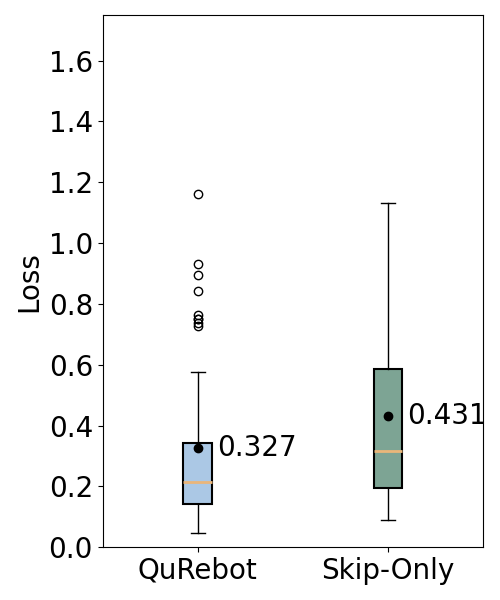}
\caption{$\FS_7$, $\horizon=2$}
\end{subfigure}
\begin{subfigure}[b]{0.16\columnwidth}
\includegraphics[width=\columnwidth]{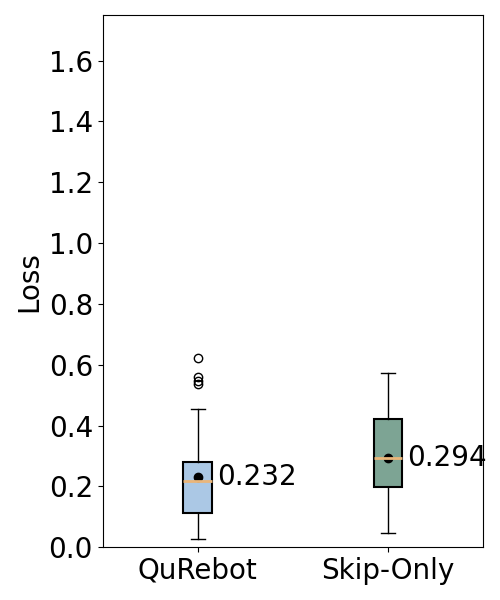}
\caption{$\FS_5$, $\horizon=2$}
\end{subfigure}
\begin{subfigure}[b]{0.16\columnwidth}
\includegraphics[width=\columnwidth]{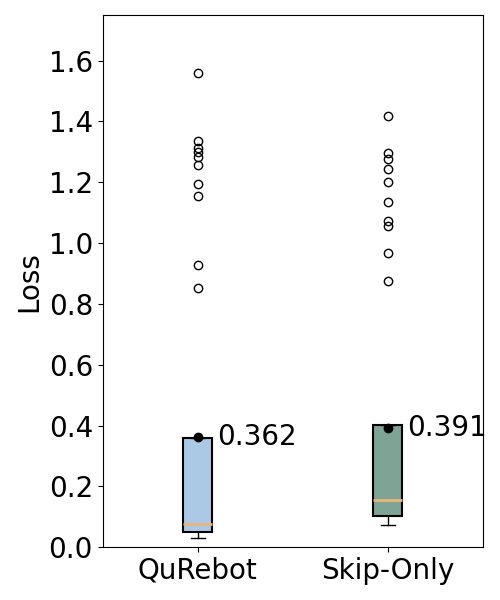}
\caption{$\FS_4$, $\horizon=2$}
\end{subfigure}

\vspace{0.5em}

\begin{subfigure}[b]{0.16\columnwidth}
\includegraphics[width=\columnwidth]{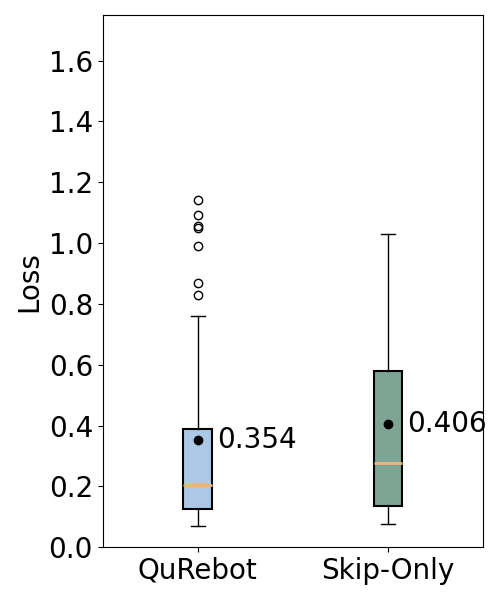}
\caption{$\FS_7$, $\horizon=3$}
\end{subfigure}
\begin{subfigure}[b]{0.16\columnwidth}
\includegraphics[width=\columnwidth]{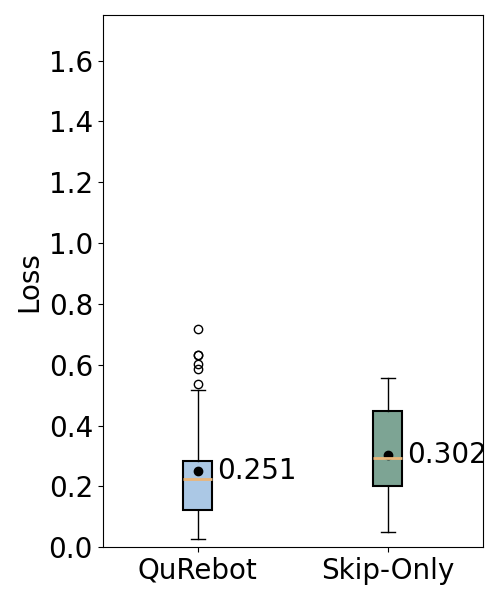}
\caption{$\FS_5$, $\horizon=3$}
\end{subfigure}
\begin{subfigure}[b]{0.16\columnwidth}
\includegraphics[width=\columnwidth]{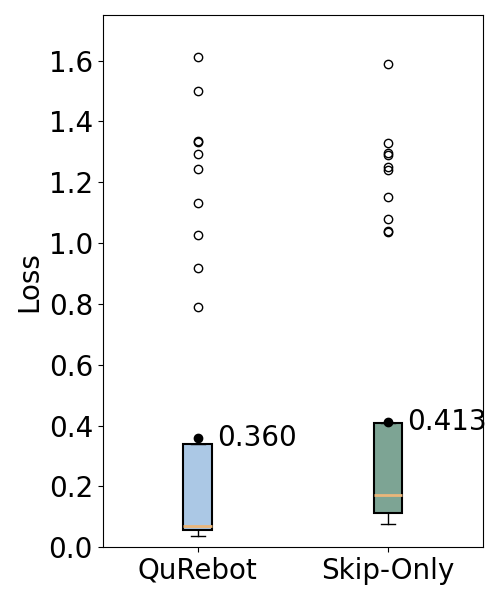}
\caption{$\FS_4$, $\horizon=3$}
\end{subfigure}

\vspace{0.5em}

\begin{subfigure}[b]{0.16\columnwidth}
\includegraphics[width=\columnwidth]{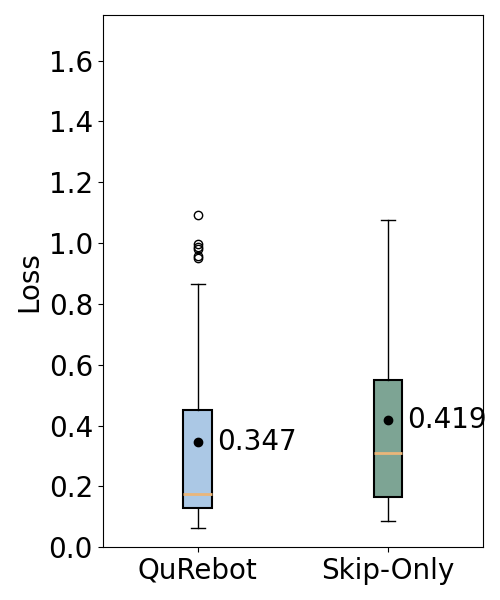}
\caption{ $\FS_7$, $\horizon=4$}
\end{subfigure}
\begin{subfigure}[b]{0.16\columnwidth}
\includegraphics[width=\columnwidth]{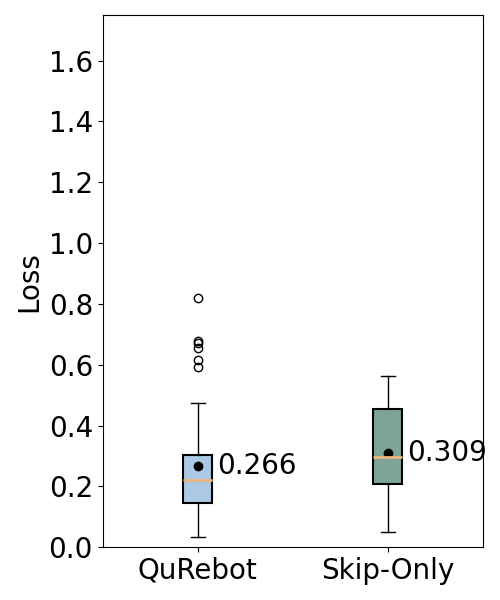}
\caption{ $\FS_5$, $\horizon=4$}
\end{subfigure}
\begin{subfigure}[b]{0.16\columnwidth}
\includegraphics[width=\columnwidth]{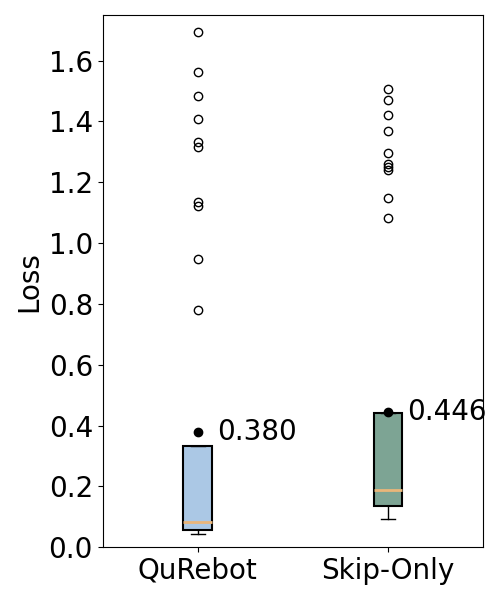}
\caption{ $\FS_4$, $\horizon=4$}
\end{subfigure}

\vspace{0.5em}

\begin{subfigure}[b]{0.16\columnwidth} 
\includegraphics[width=\columnwidth]{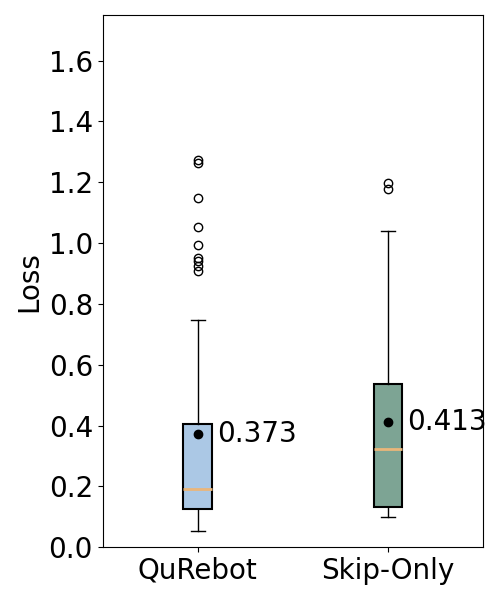}
\caption{$\FS_7$, $\horizon=5$}
\end{subfigure}
\begin{subfigure}[b]{0.16\columnwidth}
\includegraphics[width=\columnwidth]{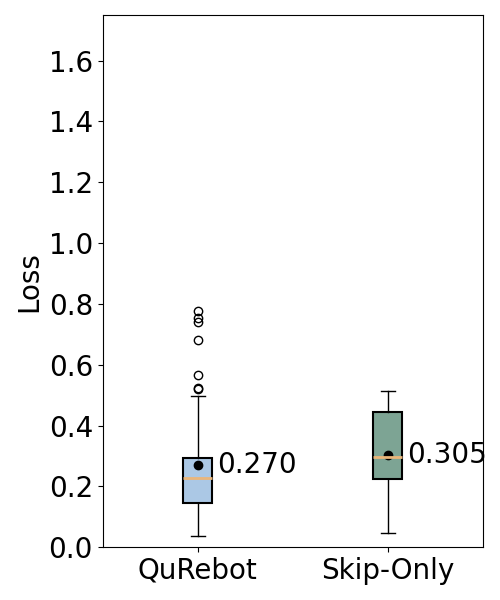}
\caption{ $\FS_5$, $\horizon=5$}
\end{subfigure}
\begin{subfigure}[b]{0.16\columnwidth}
\includegraphics[width=\columnwidth]{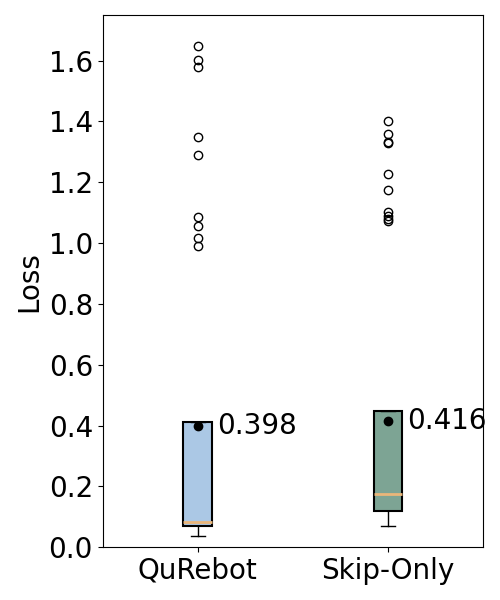}
\caption{$\FS_4$, $\horizon=5$}
\end{subfigure}
\caption{RQ1 -- \LMSE of \ourApproach and the \Skiponly model across different feature sets and horizons. The black dot is the average value of \LMSE in each box.}
\label{fig:rq1_compare}
\end{figure}

During training, all configurations of both \ourApproach and the \Skiponly model converged within 500 epochs. As Fig.~\ref{fig:rq1_compare} shows, the average \LMSE of \ourApproach is consistently lower than \Skiponly, and its boxplots are generally positioned below those of \Skiponly.
Furthermore, we conduct statistical tests on the two models for each configuration (see Table~\ref{tab:rq1_stats}).
\begin{table}[!tb]
\centering
\caption{RQ1 -- Comparison of \ourApproach and \Skiponly model with various feature sets and horizons.}
\label{tab:rq1_stats}
\renewcommand{\arraystretch}{1.4}
\resizebox{0.8\columnwidth}{!}{
\begin{tabular}{ccccccc}
\toprule
\textbf{Horizon (\horizon)} 
& \multicolumn{2}{c}{$\FS_7$} 
& \multicolumn{2}{c}{$\FS_5$} 
& \multicolumn{2}{c}{$\FS_4$} \\
\cmidrule(lr){2-3} \cmidrule(lr){4-5} \cmidrule(lr){6-7}
& $p$-value & \Atwelve & $p$-value & \Atwelve & $p$-value & \Atwelve \\
\midrule
$\horizon=1$ & 0.0491 & 0.3719 & 0.0017 & 0.2963 & 0.0078 & 0.3269 \\

$\horizon=2$ & 0.0480 & 0.3713 & 0.0037 & 0.3113 & 0.0119 & 0.3362 \\

$\horizon=3$ & 0.1673 & --- & 0.0007 & 0.2787 & 0.0229 & 0.3519 \\

$\horizon=4$ & 0.0331 & 0.3613 & 0.0004 & 0.2694 & 0.0315 & 0.3600 \\

$\horizon=5$ & 0.1673 & --- & 0.0006 & 0.2781 & 0.0331 & 0.3613 \\
\bottomrule
\end{tabular}
}
\end{table}
In most cases, the $p$-values are below 0.05 and \Atwelve is less than 0.5, indicating that \ourApproach significantly outperforms \Skiponly. The only exceptions are when using the full feature set ($\FS_7$) at $\horizon=3$ and $\horizon=5$, where the differences are not statistically significant.


These results suggest that, with the same input, the QRC branch enhances prediction accuracy in multi-step prediction tasks. Notably, \Atwelve values for $\FS_7$ at $\horizon=1$, $\horizon=2$, and $\horizon=4$ are higher than in configurations of other feature sets (except for $\FS_4$ at $\horizon=5$), indicating a larger performance gap between \ourApproach and \Skiponly when fewer features are available. This likely reflects the QRC's ability to compensate for reduced input dimensionality with quantum dynamics to extract richer patterns, mapping features into a higher-dimensional space, and supporting robust predictions~\cite{mujal2021opportunities, gotting2023exploring}.


Finally, the \LMSE of \ourApproach (see Fig.\ref{fig:rq1_compare}) shows that the Skip connection branch significantly improves prediction effectiveness of \ourApproach compared to \QRConly, enabling stable convergence and achieving lower loss values.
Across all configurations, \ourApproach achieves, on average, a 15\% improvement in prediction effectiveness over the \Skiponly. It shows that quantum dynamics in QRC indeed helps in extracting patterns from features and enhancing prediction effectiveness.

\begin{tcolorbox}[size=title, colframe=white, width=1\linewidth,
breakable,
colback=gray!20]
\textbf{Answer to RQ1}:
The skip connection branch enhances prediction accuracy over \QRConly. Also, \ourApproach outperforms \Skiponly in most configurations, highlighting the QRC branch’s ability to capture meaningful patterns. These results indicate that both branches contribute to the improved performance of \ourApproach. 
\end{tcolorbox}

\subsection{RQ2 -- Performance of \ourApproach across different configurations} \label{sec:rq2}
We compare \ourApproach across its different configurations (feature sets \FS and horizons \horizon) to help selecting an optimal configuration given testing requirements.

\begin{figure}[!tb]
\centering
\begin{subfigure}[b]{0.4\columnwidth}
\includegraphics[width=\columnwidth]{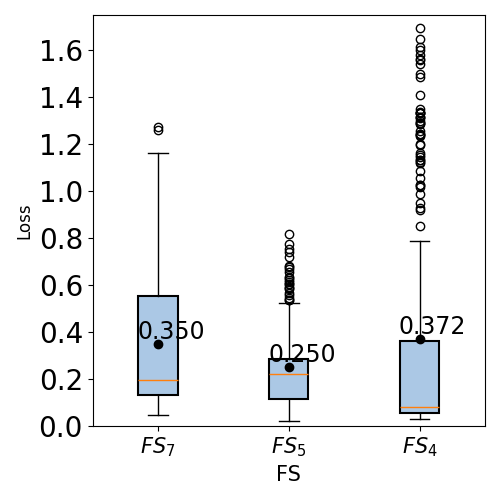}
\caption{Performance across \FS}
\label{fig:rq2_compare_fs}
\end{subfigure}
\begin{subfigure}[b]{0.4\columnwidth}
\includegraphics[width=\columnwidth]{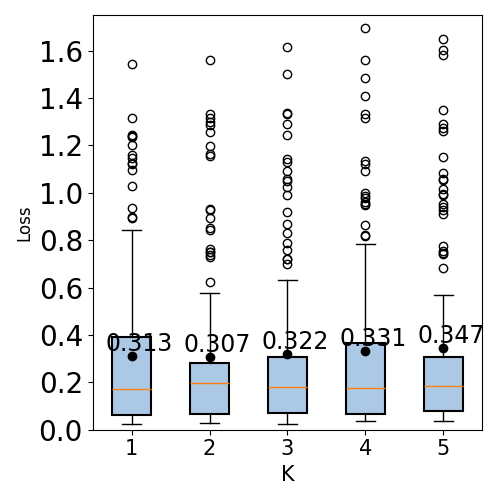}
\caption{Performance across \horizon}
\label{fig:rq2_compare_k}
\end{subfigure}
\caption{RQ2 -- \LMSE of \ourApproach with various feature sets and horizons.}
\end{figure}

First, we compare the performance of \ourApproach with different feature sets \FS. In Fig.~\ref{fig:rq2_compare_fs}, each box contains \ourApproach with the same feature set across varying prediction horizons.
Comparing the performance of the three boxes, we observe that, without outliers, \ourApproach with $\FS_7$ displays the widest range, with the highest maximum value. $\FS_5$ achieves the lowest average \LMSE and the smallest variation.
We then conduct the Friedman test on the three feature sets. Results show that the $p$-value is smaller than 0.05, indicating a significant difference among their performance. Next, we perform a pairwise Wilcoxon signed rank test and \Atwelve effect size; we find that both \ourApproach with $\FS_4$ and $\FS_5$ significantly outperforms that with $\FS_7$, while there is no significant difference between $\FS_4$ and $\FS_5$. These results suggest that \ourApproach improves the prediction effectiveness even with fewer features. A potential reason is that the data related to the robot's velocity and orientation may introduce interference, negatively affecting the performance when predicting the next positions. Thus, when predicting next states with \ourApproach for similar datasets, one can choose one feature set (either with orientation or with velocity).

Next, we investigate the performance of \ourApproach across varying prediction horizons \horizon. Fig.~\ref{fig:rq2_compare_k} depicts the general performance of \ourApproach grouped by horizon. Each box contains \ourApproach with different feature sets at a fixed horizon. We also conduct statistical tests among the five horizons. The Friedman test reveals a significant difference in performance ($p < 0.05$) among them. The results of the Wilcoxon signed rank test are shown in Table~\ref{tab:wilcoxon}.
\begin{table}[!tb]
\centering
\caption{RQ2 -- Wilcoxon signed rank test among \horizon (\textgreater\ indicates row is significantly better than column; \textless\ indicates worse; -- means not significant).}
\label{tab:wilcoxon}
\resizebox{0.7\columnwidth}{!}{
\begin{tabular}{lccccc}
\toprule
& $\horizon=1$ & $\horizon=2$ & $\horizon=3$ & $\horizon=4$ & $\horizon=5$ \\
\midrule
$\horizon=1$ & -- & -- & -- & $>$ & $>$ \\
$\horizon=2$ & -- & -- & -- & -- & $>$ \\
$\horizon=3$ & -- & -- & -- & -- & $>$ \\
$\horizon=4$ & $<$ & -- & -- & -- & -- \\
$\horizon=5$ & $<$ & $<$ & $<$ & -- & -- \\
\bottomrule
\end{tabular}
}
\end{table}
We can observe that \ourApproach with $\horizon=1$, $\horizon=2$, and $\FS_3$ consistently outperforms that with $\horizon=5$, while there is no significant difference among $\horizon=1$, $\horizon=2$, and $\horizon=3$. Also, \ourApproach with $\horizon=1$ is significantly better than $\horizon=4$, but $\horizon=4$ does not differ significantly from the remaining configurations, placing it in an intermediate position. These findings are supported by Fig.~\ref{fig:rq2_compare_k}, as the average values of $\horizon=1$, $\horizon=2$, and $\horizon=3$ are lower than the others, and a small increasing trend can be observed with the increase in horizon length. Those results indicate that \ourApproach achieves reliable performance for short-term predictions (up to 3 steps), but prediction effectiveness may degrade for longer horizons, particularly beyond 3 steps.


\begin{tcolorbox}[size=title, colframe=white, width=1\linewidth,
breakable,
colback=gray!20]
\textbf{Answer to RQ2}:
Using fewer, well-chosen features and shorter prediction horizons improves the effectiveness and reliability of \ourApproach for robotic testing.
\end{tcolorbox}

\section{Lessons Learned}
\subsection{Importance of feature selection in practice}
RQ1 (Sect.~\ref{sec:rq1}) reveals that even with a reduced number of input features, quantum dynamics can help maintain high prediction effectiveness. Moreover, RQ2 (Sect.~\ref{sec:rq2}) reveals that using more features in \ourApproach does not necessarily lead to better performance. These results indicate the importance of optimal feature selection for maximizing model performance. Moreover, using fewer features reduces the required number of qubits for the QRC circuit, resulting in lower-depth quantum circuits, lowering QC computational costs during training and testing. \looseness=-1 

\subsection{Robustness of \ourApproach}
We performed 4-fold cross-validation by dividing the dataset into four equal phases based on the time sequence of a complete robot operation. These segments correspond to a starting phase, two task execution phases, and a finishing phase, each exhibiting different characteristics in robot states. Despite the variation across phases, the results show that \ourApproach maintains consistent performance, with MSE values remaining below 1.8 across all configurations, which is within an acceptable range. It also exhibits the robustness of \ourApproach for different scenarios. More detailed analysis of \ourApproach across different phases is planned as our future work, to identify which phases in our approach can lead to relatively better or worse performance.

\subsection{Practical implication}
Our results showed that \ourApproach, a quantum-classical machine learning model, is promising for improving regression testing of AMRs using the strengths of both quantum and classical components. For PAL Robotics, our findings have several implications. First, it provides early but practical insights into the use of QML to support the AMR development. Second, our work serves as an initial step toward developing the necessary skills for applying QML in real-world robotic contexts. Third, it opens the door for PAL Robotics to explore a wider range of tasks that can benefit from QML in its practice. \looseness=-1






\section{Threats to validity}\label{sec:threats}
\textit{Internal Validity.} \ourApproach adopts a non-linear transformation as the residual connection branch. Using alternative ML models (e.g., RNN and Convolutional Neural Network) can potentially influence \ourApproach's performance. However, using more advanced ML approaches for the residual connection branch might incur more resource overhead, contradicting its role as an alternative ``shortcut'' for QRC. Nonetheless, we plan to perform more comprehensive investigations on the influence of different ML methods.

\textit{Conclusion Validity.} The inherent randomness of the neural networks and QRC can potentially influence our experimental results. To alleviate its influence and provide reliable results, we repeated each experiment 10 times and conducted statistical testing to assess the significance of differences.

\textit{External Validity.} \ourApproach is evaluated on one subject system---TIAGo OMNI robot. The performance of \ourApproach might vary for different robots. However, \ourApproach is a data-driven, hardware-agnostic approach that can be applied to most systems producing similar data structures. Moreover, we collected data from a complex and representative scenario (i.e., in a real office), where robots typically operate. We will further study the generalizability of \ourApproach and investigate its effectiveness in various scenarios in the future.

\section{Related Work}\label{sec:related}
Robotic systems' characteristics, such as autonomy, frequent interaction with the environment, including humans, make their testing critical~\cite{afzal2020study}. Several testing techniques are adopted for robots~\cite{araujo2023testing}, such as model-based testing~\cite{kanter2020model, lindvall2017metamorphic}, simulation-based testing~\cite{lu2024epitester, timperley2018crashing, sotiropoulos2017can}, formal verification~\cite{zhao2019towards, webster2020corroborative}, and runtime monitoring~\cite{huang2014rosrv,desai2017combining}. Regression testing ensures that updates or changes in the robotic software do not negatively affect the existing functionality of the robot, and also ensures new functionality is correct~\cite{afzal2020study, wienke2017performance, honfi2017model}. This paper proposes a hybrid quantum machine learning algorithm to support regression testing of an AMR from PAL Robotics. It is the first exploration of applying quantum algorithms to test an industrial robot in a real context.

Studies have used ML algorithms related to test oracles, including extracting test verdicts~\cite{braga2018machine, gholami2018classifier}, identifying metamorphic relations~\cite{hardin2018using, zhang2017rbf, hiremath2020automated}, and predicting expected outputs~\cite{monsefi2019performing, zhang2018automatic, braga2018machine, gartziandia2022machine, arrieta2021using}. Predicting expected output is related to our work, as we generate the expected behavior of the robots as the test oracle. A related study uses quantum extreme learning machine (QELM) to generate a test oracle for elevators~\cite{wang2024application}. Our work differs as we target robots, where we collect robot states during its operation and predict the future states. This involves processing temporal information, for which QELM is not designed, since it is memoryless.

QRC has gained interest~\cite{fujii2017harnessing, PhysRevResearch.3.013077, martinez2020information, domingo2022optimal, gotting2023exploring} due to its inherent ability to process temporal information based on the nature of quantum dynamics. QRC has been applied to some classical tasks, including the NARMA benchmark task~\cite{nakajima2019boosting}, an object classification task~\cite{QRC}, a quantum chemistry problem~\cite{domingo2023taking}, gene regulatory networks~\cite{xia2023configured}, and mobile user trajectory prediction~\cite{mlika2023user}. Our work differs from these works in these ways: (1) we address next state prediction for a robot, including multi-step prediction. Also, previous studies usually focus on single-variable signals, whereas our input is multi-dimensional state information (e.g., position, velocity) as well as we predict more than one signal, which makes our task more difficult. (2) we propose a hybrid framework that combines both QRC and a ``skip connection'' inspired by residual connection in deep learning, significantly enhancing the QRC performance.

\section{Conclusion and future work}
This paper generated an ML-based test oracle for the TIAGo OMNI robot developed by \textit{PAL Robotics}. Particularly, we propose a hybrid framework, \ourApproach, that combines the QRC algorithm and a neural network. 
Specifically, we use this framework for the next state prediction of the robot. We evaluate \ourApproach with the navigation dataset of the robot in an office map. We compare \ourApproach with a QRC baseline and a neural network baseline. Results show that \ourApproach outperforms both models. In addition, we compare the performance of \ourApproach under various configurations (i.e., input feature sets and prediction horizons). Then, we provide practical guidance on optimal configurations with higher prediction effectiveness and discuss practical implications. In the future, we will implement \ourApproach on real quantum computers to evaluate the effect of quantum noise as well as evaluate its robustness in other industrial and realistic contexts.

\bibliographystyle{unsrt}  
\bibliography{biblio}  


\end{document}